\documentclass[conference]{IEEEtran}
\IEEEoverridecommandlockouts
\usepackage{cite}
\usepackage{multirow}
\usepackage{amsmath,amssymb,amsfonts}
\usepackage{algorithmic}
\usepackage{graphicx}
\usepackage{textcomp}
\usepackage{xcolor}
\usepackage{booktabs}
\usepackage{longtable}
\usepackage{caption}
\usepackage{subcaption}
\usepackage{tikz}
\usepackage{units}
\usetikzlibrary{positioning}

\def\BibTeX{{\rm B\kern-.05em{\sc i\kern-.025em b}\kern-.08em
    T\kern-.1667em\lower.7ex\hbox{E}\kern-.125emX}}

\usepackage{flushend}
\usepackage{hyperref}
\usepackage{tikz}

\usepackage{xcolor}

\usepackage{lipsum}
\usepackage[pscoord]{eso-pic}
\newcommand{\placetextbox}[3]{
  \setbox0=\hbox{#3}
  \AddToShipoutPictureFG*{
    \put(\LenToUnit{#1\paperwidth},\LenToUnit{#2\paperheight}){\vtop{{\null}\makebox[0pt][c]{#3}}}%
  }%
}%

\begin{document}
\placetextbox{0.5}{1}{This is the author's version of an article that has been published.}
\placetextbox{0.5}{0.985}{Changes were made to this version by the publisher prior to publication.}
\placetextbox{0.5}{0.97}{The final version of record is available at \href{https://doi.org/10.1109/ISIE62713.2025.11124605}{https://doi.org/10.1109/ISIE62713.2025.11124605}}%
\placetextbox{0.5}{0.05}{Copyright (c) 2025 IEEE. Personal use is permitted.}
\placetextbox{0.5}{0.035}{For any other purposes, permission must be obtained from the IEEE by emailing pubs-permissions@ieee.org.}%

\title{Improving Wi-Fi Network Performance Prediction with Deep Learning Models\thanks{This work was partially supported by the European Union under the Italian National Recovery and Resilience Plan (NRRP) of NextGenerationEU, partnership on ``Telecommunications of the Future'' (PE00000001 - program ``RESTART'') as well as the Swedish Knowledge Foundation through the research profile NIIT.}}

\author{
\IEEEauthorblockN{Gabriele~Formis\IEEEauthorrefmark{1}\IEEEauthorrefmark{3}, Amanda~Ericson\IEEEauthorrefmark{2}, Stefan~Forsström\IEEEauthorrefmark{2}, Kyi~Thar\IEEEauthorrefmark{2}, Gianluca~Cena\IEEEauthorrefmark{1}, and  Stefano~Scanzio\IEEEauthorrefmark{1}}
\IEEEauthorblockA{\IEEEauthorrefmark{1}\textit{National Research Council of Italy (CNR–IEIIT), Italy} \IEEEauthorrefmark{3}\textit{Politecnico di Torino, Italy}\\}
\IEEEauthorblockA{\IEEEauthorrefmark{2}\textit{Department of Computer and Electrical Engineering, Mid Sweden University, Sundsvall, Sweden}\\
gabrieleformis@cnr.it, \{amanda.ericson, stefan.forsstrom,  kyi.thar\}@miun.se,\{gianluca.cena, stefano.scanzio\}@cnr.it
}
}

\maketitle

\begin{abstract}
The increasing need for robustness, reliability, and determinism in wireless networks for industrial and mission-critical applications is the driver for the growth of new innovative methods. The study presented in this work makes use of machine learning techniques to predict channel quality in a Wi-Fi network in terms of the frame delivery ratio. Predictions can be used proactively to adjust communication parameters at runtime and optimize network operations for industrial applications. Methods including convolutional neural networks and long short-term memory were analyzed on datasets acquired from a real \mbox{Wi-Fi} setup across multiple channels. The models were compared in terms of prediction accuracy and computational complexity. Results show that the frame delivery ratio can be reliably predicted, and convolutional neural networks, although slightly less effective than other models, are more efficient in terms of CPU usage and memory consumption. This enhances the model's usability on embedded and industrial systems.
\end{abstract}


\section{Introduction}
Robustness and dependability are the main challenges in next-generation communication systems, especially in wireless networks for industrial applications like Wi-Fi \cite{10372393}, but also in the context of smart cities and buildings, transportation, and agriculture. Factory automation and real-time monitoring require communication technologies that can maintain predictable performance even under fluctuating conditions of the quality of the channel \cite{10317890}. 
Due to the numerous sources of interference and electromagnetic noise,
such as machinery, and the coexistence of multiple wireless systems \cite{2021103388},
industrial environments are particularly prone to these fluctuations. 
Making a wireless connection able to fulfill the requirements of industrial applications,
in terms of determinism, resilience, and bounded latencies, 
in these harsh conditions makes the automatic tuning of communication parameters to channel quality variations truly challenging.

Several methods are available in the scientific literature to improve the quality of a Wi-Fi link,
some of which
exploit the prediction of the future quality of the wireless channel (which is the target of this work) to proactively react to its variations, in order to maintain stable performance indicators.
The dynamic and complex nature of the wireless spectrum makes the adoption of traditional probabilistic approaches based on statistics not so effective. 
The perceived quality of wireless communication depends on factors such as the number and type of interferers, the communication technologies used, e.g., Wi-Fi (IEEE 802.11), Bluetooth (IEEE 802.15.1), low-rate wireless personal area networks (LR-WPANs, IEEE 802.15.4), etc., the operating environment, the mobility of nodes, and so on. 
Taken together, all of these effects make reliable channel prediction a very complex task and a significant research area with many practical implications. 
Among the different key performance indicators to be predicted, this work focuses on the frame delivery ratio (FDR).
This quantity can be used by real-time applications to directly infer the future quality of the wireless channel and to timely trigger suitable corrective actions. 
For instance, knowing what the FDR will be in the near future can be exploited to guarantee real-time constraints.   

Generally speaking, machine learning (ML) algorithms can be used to capture temporal dependencies. 
To address the related challenges, advanced ML models such as convolutional neural networks (CNN), long short-term memory (LSTM), and bidirectional LSTM (Bi-LSTM) are investigated in this work for the first time in the context of Wi-Fi channel quality prediction in industrial scenarios. 
These models are particularly effective in analyzing sequential data, making them well-suited for predicting the FDR in such contexts. 
By comparing their performance, this work highlights the strengths and limitations of every single approach, focusing on its ability to handle varying network conditions and to balance computational efficiency with prediction accuracy.

The estimation of the FDR performed by the ML models proposed and compared here can be profitably used to change specific network and application parameters. 
When the predicted FDR worsens, the application could, e.g., decrease the sampling rate or the allowed maximum amount of concurrent best-effort traffic, to reduce the overall load on the wireless channel and consequently improve its real-time quality. 
Other important contexts where the prediction of the FDR can be used in practice include the selection at runtime of the best channel for redundant transmission schemes \cite{8502636}, the automatic configuration of network parameters \cite{9779183}, and enhanced handover procedures that may anticipate some operations in order to perform an early channel switching \cite{Khan2022}. 
Finally, forecasts can also be exploited to improve the effectiveness of Wi-Fi network digital twins \cite{10540718}. 
Suitable algorithms for implementing above
mechanisms could be integrated into the wireless adapter driver.

The remainder of this paper is {organized} as follows.
Section~\ref{sec-rw} provides an overview of selected related work on wireless channel prediction, with a focus on deep learning models such as CNN, LSTM, and other time series prediction techniques.
Section~\ref{sec-method} describes the methodology we used for dataset acquisition, data preprocessing, and model selection.
Section \ref{sec-implementation} presents the implementation details of the analyzed ML models, including their architectural design, the selection of hyperparameters, and how hyperparameters are tuned,
while
Section~\ref{sec-result} reports comparative results 
concerning
the accuracy of CNNs, 
LSTM, and Bi-LSTM networks, as well as their computational complexity.
Section~\ref{sec-discussion} discusses the implications of 
adopting the proposed models in real-world applications that rely on industrial wireless networks, and finally Section~\ref{sec-conclusion} concludes the paper by summarizing the key achievements and suggesting future research directions.

\begin{figure*}[t]
\includegraphics[width=1\linewidth]{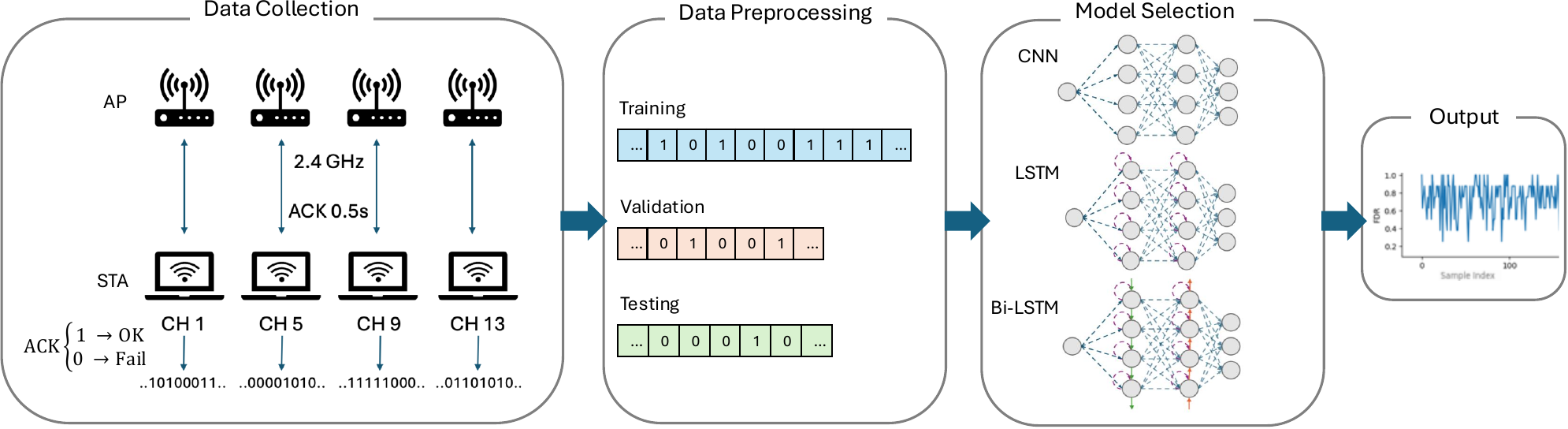}
\centering
\caption{Main steps of the proposed methodology for predicting Wi-Fi network performance.}
\label{fig:pipeline}
\centering
\end{figure*}
\section{Related Work} \label{sec-rw}
The increasing demand for dependability in wireless communications has led to the proposal of various channel quality prediction models.  
In particular, the application of ML to \mbox{Wi-Fi} has been a constantly growing research trend \cite{10726906,9786784},
and the use of ML models 
has been suggested
in several contexts concerning industrial communication,
a concrete example being handover management \cite{10710709}.

Due to their simplicity and computational efficiency,
traditional statistical approaches \cite{10144122,5288559} are still widely used 
for providing satisfactory forecasts about the quality of a wireless link.
Studies have shown that models based on exponential moving averages (EMA) 
can effectively predict the FDR based on historical data, offering a cost-effective alternative to complex machine learning models. 
The linear combination of EMA models (ELC)\cite{formis2023linear} 
has also been proposed to improve prediction accuracy over traditional EMA models, 
with a slight increase in computational complexity.

More recently, common ML models such as artificial neural networks (ANN) of multilayer perceptron (MLP) type \cite{scanzio2022predicting} have gained attention for channel quality prediction. 
Interestingly, models like the deep learning encoder-decoder architecture have shown significantly better performance than statistical methods \cite{herath2019deep}, but not for FDR estimation.

LSTM networks \cite{kanto2024wireless} have been recognized as an effective tool for predicting the behavior of wireless channels in the case of sequential data \cite{stenhammar2024comparison}. 
Comparative studies indicate that LSTM-based models outperform conventional methods, such as auto-regression and linear regression, in the prediction of the signal strength for different communication standards, including Wi-Fi and 4G networks \cite{herath2019deep}.

Recent research has also focused on comparing the performance of different ANN architectures, including MLP, CNN, and transformers, for channel prediction \cite{jiang2022accurate}. These comparisons demonstrate that while models like LSTM and gated recurrent unit (GRU) work well to capture sequential dependencies, other architectures such as CNNs can be optimized to recognize spatio-temporal patterns in multi-antenna configurations \cite{liu2014temporal}.
Although ANN-based models with a large number of parameters provide higher accuracy, they require more computational resources, which can limit their applicability in embedded systems. Hybrid approaches that combine traditional statistical models with machine learning offer a trade-off between accuracy and computational efficiency \cite{ericson2024iiot}.

\section{Method} \label{sec-method}
An overview of the proposed methodologies behind the prediction of the FDR is outlined in Fig. \ref{fig:pipeline}. It covers some key stages, including data collection, data processing, and model selection. 
These stages are aimed at finding a trained model that is usable to predict the FDR without requiring excessive resources in terms of computational power and memory footprint.

\subsection{Data collection} 
The collected dataset contains more than $186$ million patterns, which are acquired on four non-overlapping Wi-Fi channels (1, 5, 9, and 13) in the \unit[2.4]{GHz} band. This band was selected because it is characterized by intense interfering traffic due to several networks located near the one used for dataset acquisition. 
The overall duration of the dataset corresponds to more than $1000$ days, i.e., more than $250$ days for every channel. 
The evolution of the FDR over time differs significantly for the four considered channels.

For dataset acquisition, many features of the Wi-Fi protocol were disabled, including frame retransmissions, rate adaptation (speed was set to \unit[54]{Mbps}), backoff, and frame aggregation. 
This is because our aim was to periodically sample the quality of the channel by performing a single acknowledged (ACK) transmission attempt every $0.5$ seconds. 
Disabling some protocol features and using the selected transmission period had the goal of impacting channel quality as few as possible, by injecting a minimal amount of additional traffic and sampling the channel at a constant rate.
The software and hardware used for dataset collection were previously described in \cite{10295470}.

Every pattern in the dataset includes features such as the transmission outcome (success/failure), receive signal strength indicator (RSSI) of the ACK frame, and transmission latency.
This work specifically focuses on transmission outcomes. The analysis of the other acquired features is left as future work.
In particular, upon the reception at discrete time $i$ of the ACK frame that is sent back by the receiver---which in our case is the access point (AP)---to notify the correct reception of the data frame, the transmission outcome $x_i=1$ (corresponding to a successful transmission) is logged. 
Conversely, in the case the ACK frame does not arrive (which could be due to the loss of either the data or the ACK frame), the outcome $x_i=0$ is logged. 
We point out that all the features that permit the prediction of the FDR are intentionally acquired on the sending node, so that relevant decisions about countermeasures can be made locally, depending on its estimated value.
The collected dataset consists of the ordered sequence of all the outcomes $(x_1, ...,x_i, ..., x_N)$
and includes $N$ patterns.

\subsection{Data Preprocessing}
The acquired set of data was subdivided into three datasets, disjoint in time. For each one of the four channels, $20\%$ of patterns were selected for testing, $20\%$ for validation, and $60\%$ for training.
The two classes of the dataset related to outcomes are quite unbalanced: $x_i=0$ corresponds to 15\% of the patterns, while $x_i=1$ corresponds to 85\% of the patterns. In this context, a simple oversampling technique was first tested, which consists of repeating the patterns related to the underbalanced class to make the cardinality of the two classes equal. However, it was not used, as it did not provide tangible improvements while greatly increasing the training time of the model.

Let $i$ be the current time instant. 
The corresponding target 
\begin{align}
    t_i=\frac{1}{N_\mathrm{f}}\sum_{j=i+1}^{i+N_\mathrm{f}}x_j
\end{align} 
(i.e., the FDR) is computed by averaging the following $N_\mathrm{f}=3600$ outcomes,
which correspond to $30$ minutes.

\subsection{Model Selection}
The process of selecting ML models focused firstly on the accuracy of prediction in terms of FDR in non-stationary network conditions, which are typical for real environments. Models were chosen to ensure a balance between simplicity, computational efficiency, and the ability to model temporal dependencies, satisfying the requirements of industrial applications. 
LSTM models were chosen for their ability to identify patterns in sequential data. 
Instead, Bi-LSTM models were considered for their ability to capture both past and future dependencies, providing a more comprehensive understanding of temporal patterns. 
Finally, CNN models were evaluated as a simpler and more efficient alternative for predicting the FDR in devices with low computational power. Performance comparisons among these models are a main goal of this work.

\section{Implementation}\label{sec-implementation}
This section focuses on the technical decisions and adaptations made while implementing the proposed technique, including architectural design and hyperparameter tuning.
All the ML models were trained with the \texttt{Keras} module of \texttt{TensorFlow}. 
To be more adherent to the implementation details and to enhance reproducibility, in the next sections the \texttt{Keras} nomenclature is directly used to describe the implemented software.

\subsection{Models}

The input of each model is a sequence of outcomes of length $l$, which is shuffled at each training epoch in the case of CNN and is provided to the ML model sequentially in the case of LSTM and Bi-LSTM.

The CNN model consists of one \texttt{Conv1D} convolutional layer, which was parameterized by the number of \texttt{filters} and the \texttt{kernel\_size}, and uses \textit{ReLu} as activation function. As usual, a \texttt{Flatten} layer is used to transform the output of the convolutional layer in a single dimension. The following layers are three \texttt{Dense} layers (four in the case of one experimental configuration). The first two (or three) are based on \textit{ReLu}, while the latter is \textit{linear} because the model was used in a regressive fashion to predict the FDR.
A max-pooling layer was not used because it causes information loss when the sequence is time-dependent. A \textit{dropout} layer, which is typically used to reduce overfitting, was tested but not used because it did not improve the prediction accuracy.

The LSTM model consists of one \texttt{LSTM} layer with \textit{hyperbolic tangent} activating function, followed by a \texttt{Dense} layer with a \textit{linear} output. 
The Bi-LSTM Model is similar to the LSTM model but includes bidirectional connections to capture dependencies in both the forward and backward directions. This architecture improves the ability of the model, compared to traditional LSTM, to handle complex temporal patterns.

The selected loss function for all models was the \textit{mean squared error} with the \textit{Adam} optimizer, which was chosen due to its robustness in managing varying gradient magnitudes.

\subsection{Hyperparameter tuning}
Hyperparameter optimization played a critical role in achieving optimal performance for each model. Different combinations of learning rate, batch size, training epochs, sequence lengths $l$, number of units, number of layers, and learning rate decay policy were explored to find the best model parameterization. 
The learning rate decay was implemented to improve convergence during training, by halving the learning rate each epoch. 
Early stopping was used in all the experimental campaigns to monitor validation loss and prevent overfitting.

To better select the optimal hyperparameters, the \texttt{fmin} function of the \texttt{hyperopt} Python library was employed, with the goal of minimizing the following objective function:
\begin{equation}
J(M, \theta) = \sum_{p_i, t_i \in 
 \mathcal{V}}\Big( t_i-f_M(p_i,\theta)\Big) ^2,
 \label{eq:loss}
\end{equation}
where $J(M, \theta)$ is the loss function based on the mean squared error provided by model $M$ (i.e., CNN, LSTM, or Bi-LSTM) with hyperparameters $\theta$, $p_i$ and $t_i$ are the input patterns and targets computed on the validation dataset $\mathcal{V}$, and $f_M(p_i,\theta)$ is the prediction of model $M$ evaluated with the input pattern $p_i$ and the model hyperparameters $\theta$.

\begin{figure}[t]
\includegraphics[width=1\linewidth]{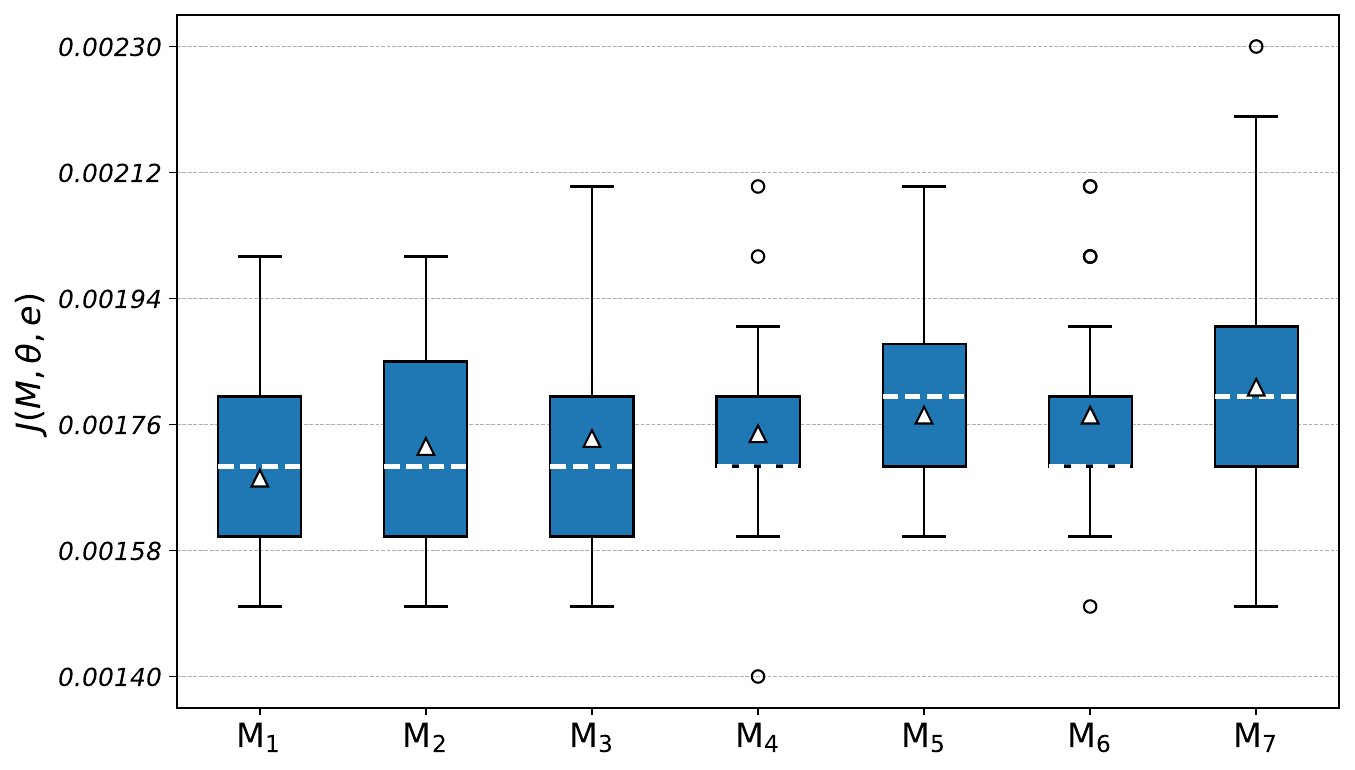}
\centering
\vspace{0.1cm}
\caption{Box plot showing the loss on the validation dataset $\mathcal{V}$ for channel 1 concerning the seven best configurations related to the CNN model.}
\label{fig:box_plot}
\centering
\end{figure}

All the hyperparameters $\theta$ examined by the \texttt{fmin} function and the quality of the models obtained for each training epoch were analyzed. Let $J(M, \theta, \tau)$ be the loss computed for model $M$, hyperparameter $\theta$, and training epoch $\tau$. All these $J(M, \theta, \tau)$ values are summarized by the box plot in Fig.~\ref{fig:box_plot} in the case of the CNN model trained and tested on the dataset acquired on channel 1. Among all the analyzed configurations, the best seven were reported in order of \textit{average loss} $\overline{J}(M, \theta)$ evaluated on the validation dataset $\mathcal{V}$.

For any model $M$ characterized by hyperparameters $\theta$, 
the average loss $\overline{J}(M, \theta)$ is computed by averaging the loss $J(M, \theta, \tau)$ over all the training epochs (excluding the first five ones): 
\begin{equation}
\overline{J}(M, \theta) = \frac{1}{N_\tau-5} \sum_{\tau=6}^{N_\tau}{J(M, \theta, \tau)},
\label{eq:avg_loss}
\end{equation}
where $N_\tau$ is the number of training epochs. 
The first five epochs are excluded because, for them, the model has not reached a stable behavior yet in terms of the loss.
From the box plot, it can be observed that model quality and stability highly depend on the choice of hyperparameters $\theta$. 
Since the model's losses $J(M, \theta, \tau)$ fluctuate significantly for the different training epochs, the selection of the best hyperparameters is based on $\overline{J}(M, \theta)$, which is marked with white triangles in each box of the plot. The dashed white line in the same plot represents the median.

\begin{table*}[t]
\centering
\caption{Model parameters for CNN, LSTM, and Bi-LSTM (``ch'' and ``all'' conditions)}
\label{tab:model_parameters_combined}
\normalsize
\tabcolsep=0.13cm
\def\arraystretch{1.2}
\begin{tabular}{|c|c|c|c|c|}
\hline
\textbf{Parameter}          & \multicolumn{2}{c|}{\textbf{CNN}}       & \multicolumn{2}{c|}{\textbf{LSTM / Bi-LSTM}} \\
                            & \textbf{ch} & \textbf{all}             & \textbf{ch} & \textbf{all}           \\ \hline
\textbf{Input Sequence Length} ($l$) & 3600       & 3600                    & 1200       & 1200                    \\ \hline
\textbf{Batch Size}          & 64          & 128                     & 32         & 64                        \\ \hline
\textbf{Epochs} ($N_\tau$)   & 30          & 40                      & 15         & 25                         \\ \hline
\textbf{Optimizer}           & Adam        & Adam                    & Adam       & Adam                        \\ \hline
\textbf{Loss Function}       & Mean Squared Error & Mean Squared Error & Mean Squared Error & Mean Squared Error \\ \hline
\textbf{Learning Rate}       & 0.01 (decay) & 0.005 (decay)          & 0.01 (decay) & 0.005 (decay)        \\ \hline
\textbf{Number of Filters/Units} & 128 (filters) & 256 (filters)      & 25 (LSTM units) & 25,50 (LSTM units) \\ \hline
\textbf{Kernel Size}         & 3           & 5                       & N/A        & N/A                      \\ \hline
\textbf{Pooling}             & MaxPooling (size 2) & MaxPooling (size 2) & N/A        & N/A                   \\ \hline
\textbf{Dense Layers}        & 3 (128, 64, 1 unit) & 4 (256, 128, 64, 1 unit) & 1 (1 unit) & 2 (8, 1 unit)     \\ \hline
\textbf{Activation Functions} & ReLU (Conv1D/Dense) & ReLU (Conv1D/Dense) & Tanh & Tanh          \\ \hline
\end{tabular}
\end{table*}

\subsection{Selected hyperparameters}

The selected best hyperparameters $\theta^*$, given a model $M$, are based on the hyperparametes $\theta$ that minimize $\overline{J}(M, \theta)$:
\begin{equation}
\theta^* = \arg \min_{\theta} \overline{J}(M, \theta).
\label{eq:opt_single_ch}
\end{equation}

The experimental evaluation discussed in the next Section~\ref{sec-result} reports the results for each individual channel (i.e., ``ch1'', ``ch5'', ``ch9'', and ``ch13'' conditions for channels 1, 5, 9, 13, respectively) and those where the individual channel datasets were concatenated (i.e., ``all'' condition). 
The latter ``all'' condition helps understanding the generalization capabilities of the proposed ML algorithm using only a single model for all the channels. In the case of individual channels, a single parameterization in terms of the selected hyperparameters was chosen for all of them. 
This led to a slight worsening of the prediction accuracy but, at the same time, to a choice of the hyperparameters that does not overfit (i.e., it is not specific to) the characteristics of the given channel.

For the case of the single channel analysis, hyperparameters $\theta^*$ are selected in order to minimize the average loss on the four channels as follows:
\begin{eqnarray}
\theta^* & = & \arg \min_{\theta} \frac{1}{4} \cdot \Big(  \overline{J}(M, \theta)_{\mathrm{ch1}} + \overline{J}(M, \theta)_{\mathrm{ch5}} + \nonumber \\
 & + & \overline{J}(M, \theta)_{\mathrm{ch9}} + \overline{J}(M, \theta)_{\mathrm{ch13}}
 \Big), \label{eq:avg_ch}
\end{eqnarray}
where $\overline{J}(M, \theta)_{\mathrm{ch1}}$ is the average loss evaluated on the validation dataset related to ``ch1'', $\overline{J}(M, \theta)_{\mathrm{ch5}}$ is the average loss evaluated on the validation dataset related to ``ch5'', etc.

Instead, for the ``all'' condition, the outcome provided by (\ref{eq:opt_single_ch}) can be directly used to select the best hyperparameters $\theta_{\mathrm{all}}^*$. In this case, the validation dataset $\mathcal{V}$ is the concatenation of the validation datasets of the four channels.
All the selected hyperparameters, which were used in the experimental campaign, are reported in Table~\ref{tab:model_parameters_combined}. As can be noticed from the table, the selected hyperparameters for every single ``ch'' condition and the ``all'' condition do not differ significantly. 
An interesting point is the use of a more complex model structure in the case of ``all'' (i.e., four dense layers in the case of CNN, and two dense layers in the case of LSTM and Bi-LSTM), to have a model with enough parameters for learning the larger amount of information contained in a database that is four times larger.

\begin{table*}[t]
  \caption{Metrics for the accuracy of the CNN model across different channels
  }
  \label{tab:metrics_cnn}
  \normalsize
  \begin{center}
    \tabcolsep=0.1cm
    \def\arraystretch{1.2}
    \begin{tabular}{ccc|ccccc|cccccc|cccc}
    Test & Train & Model & $\mu_{e^2}$ & $e^2_{\mathrm{p}_{90}}$ & $e^2_{\mathrm{p}_{95}}$ & $e^2_{\mathrm{p}_{99}}$ & $e^2_{\mathrm{max}}$ & $\mu_{|e|}$ & $\sigma_{|e|}$ & ${|e|}_{\mathrm{p}_{90}}$ & ${|e|}_{\mathrm{p}_{95}}$ & ${|e|}_{\mathrm{p}_{99}}$ & ${|e|}_{\mathrm{max}}$ & ${e}_{\mathrm{min}}$ & ${e}_{\mathrm{p}_{5}}$ & ${e}_{\mathrm{p}_{95}}$ & ${e}_{\mathrm{max}}$ \\
    Ch. & Ch. & & \multicolumn{5}{c|}{$[\cdot 10^{-3}]$} & \multicolumn{6}{c|}{[\%]} & \multicolumn{4}{c}{[\%]} \\
    \hline
    $1$ & $1$ & CNN & 2.15 & 5.64 & 11.77 & 30.26 & 100.24 & 2.80 & 3.69 & 7.51 & 10.85 & 17.40 & 31.66 & -31.63 & -6.41 & 4.69 & 31.66 \\
    $1$ & All & CNN & 2.03 & 5.53 & 11.62 & 29.72 & 99.83 & 2.73 & 3.63 & 7.47 & 10.79 & 17.28 & 31.48 & -31.54 & -6.34 & 4.63 & 31.44 \\
    \hline
    $5$ & $5$ & CNN & 1.17 & 1.58 & 2.70 & 17.56 & 176.41 & 2.18 & 2.64 & 3.97 & 5.20 & 13.25 & 42.00 & -42.00 & -5.20 & 3.61 & 42.00 \\
    $5$ & All & CNN & 1.13 & 1.54 & 2.67 & 17.43 & 175.88 & 2.15 & 2.62 & 3.93 & 5.16 & 13.19 & 41.83 & -41.89 & -5.17 & 3.57 & 41.73 \\
    \hline
    $9$ & $9$ & CNN & 4.23 & 5.92 & 7.81 & 60.43 & 248.19 & 4.89 & 4.29 & 7.69 & 8.84 & 24.58 & 49.82 & -49.82 & -2.10 & 6.01 & 49.82 \\
    $9$ & All & CNN & 4.11 & 5.86 & 7.74 & 60.21 & 247.47 & 4.83 & 4.25 & 7.63 & 8.79 & 24.55 & 49.78 & -49.85 & -2.09 & 5.97 & 49.77 \\
    \hline
    $13$ & $13$ & CNN & 4.89 & 9.88 & 14.61 & 22.94 & 28.45 & 6.10 & 3.42 & 9.94 & 12.09 & 15.14 & 16.87 & -13.27 & -9.52 & 11.98 & 16.87 \\
    $13$ & All & CNN & 4.73 & 9.73 & 14.49 & 22.79 & 28.33 & 6.04 & 3.37 & 9.87 & 12.06 & 15.08 & 16.74 & -13.22 & -9.47 & 11.87 & 16.72 \\
    \hline
    \end{tabular}
    \end{center}
\end{table*}

\begin{table*}[t]
  \caption{Metrics for the accuracy of the LSTM model across different channels
  }
  \label{tab:metrics_lstm}
  \normalsize
  \begin{center}
    \tabcolsep=0.1cm
    \def\arraystretch{1.2}
    \begin{tabular}{ccc|ccccc|cccccc|cccc}
    Test & Train & Model & $\mu_{e^2}$ & $e^2_{\mathrm{p}_{90}}$ & $e^2_{\mathrm{p}_{95}}$ & $e^2_{\mathrm{p}_{99}}$ & $e^2_{\mathrm{max}}$ & $\mu_{|e|}$ & $\sigma_{|e|}$ & ${|e|}_{\mathrm{p}_{90}}$ & ${|e|}_{\mathrm{p}_{95}}$ & ${|e|}_{\mathrm{p}_{99}}$ & ${|e|}_{\mathrm{max}}$ & ${e}_{\mathrm{min}}$ & ${e}_{\mathrm{p}_{5}}$ & ${e}_{\mathrm{p}_{95}}$ & ${e}_{\mathrm{max}}$ \\
    Ch. & Ch. & & \multicolumn{5}{c|}{$[\cdot 10^{-3}]$} & \multicolumn{6}{c|}{[\%]} & \multicolumn{4}{c}{[\%]} \\
    \hline
    $1$ & $1$ & LSTM & 2.08 & 5.59 & 11.68 & 30.00 & 99.90 & 2.77 & 3.67 & 7.48 & 10.82 & 17.35 & 31.50 & -31.60 & -6.39 & 4.67 & 31.50 \\
    $1$ & All & LSTM & 1.98 & 5.45 & 11.50 & 29.60 & 99.50 & 2.68 & 3.60 & 7.42 & 10.70 & 17.20 & 31.40 & -31.50 & -6.30 & 4.60 & 31.30 \\
    \hline
    $5$ & $5$ & LSTM & 1.15 & 1.57 & 2.69 & 17.50 & 176.30 & 2.16 & 2.63 & 3.95 & 5.18 & 13.23 & 41.95 & -42.10 & -5.21 & 3.60 & 42.10 \\
    $5$ & All & LSTM & 1.10 & 1.51 & 2.63 & 17.35 & 175.75 & 2.10 & 2.61 & 3.90 & 5.12 & 13.12 & 41.80 & -41.85 & -5.15 & 3.55 & 41.70 \\
    \hline
    $9$ & $9$ & LSTM & 4.21 & 5.91 & 7.80 & 60.40 & 247.90 & 4.88 & 4.27 & 7.68 & 8.85 & 24.60 & 49.90 & -49.95 & -2.12 & 6.00 & 49.95 \\
    $9$ & All & LSTM & 4.12 & 5.85 & 7.72 & 60.18 & 247.40 & 4.82 & 4.24 & 7.62 & 8.80 & 24.50 & 49.80 & -49.88 & -2.08 & 5.96 & 49.85 \\
    \hline
    $13$ & $13$ & LSTM & 4.87 & 9.87 & 14.60 & 22.90 & 28.42 & 6.11 & 3.42 & 9.92 & 12.08 & 15.12 & 16.84 & -13.25 & -9.53 & 11.95 & 16.84 \\
    $13$ & All & LSTM & 4.75 & 9.75 & 14.50 & 22.82 & 28.30 & 6.05 & 3.38 & 9.88 & 12.05 & 15.05 & 16.75 & -13.20 & -9.48 & 11.85 & 16.78 \\
    \hline
    \end{tabular}
    \end{center}
\end{table*}

\begin{table*}[t]
  \caption{Metrics for the accuracy of the Bi-LSTM model across different channels
  }
  \label{tab:metrics_bilstm}
  \normalsize
  \begin{center}
    \tabcolsep=0.1cm
    \def\arraystretch{1.2}
    \begin{tabular}{ccc|ccccc|cccccc|cccc}
    Test & Train & Model & $\mu_{e^2}$ & $e^2_{\mathrm{p}_{90}}$ & $e^2_{\mathrm{p}_{95}}$ & $e^2_{\mathrm{p}_{99}}$ & $e^2_{\mathrm{max}}$ & $\mu_{|e|}$ & $\sigma_{|e|}$ & ${|e|}_{\mathrm{p}_{90}}$ & ${|e|}_{\mathrm{p}_{95}}$ & ${|e|}_{\mathrm{p}_{99}}$ & ${|e|}_{\mathrm{max}}$ & ${e}_{\mathrm{min}}$ & ${e}_{\mathrm{p}_{5}}$ & ${e}_{\mathrm{p}_{95}}$ & ${e}_{\mathrm{max}}$ \\
    Ch. & Ch. & & \multicolumn{5}{c|}{$[\cdot 10^{-3}]$} & \multicolumn{6}{c|}{[\%]} & \multicolumn{4}{c}{[\%]} \\
    \hline
    $1$ & $1$ & Bi-LSTM & 2.07 & 5.57 & 11.65 & 29.95 & 99.85 & 2.76 & 3.66 & 7.46 & 10.80 & 17.30 & 31.45 & -31.55 & -6.37 & 4.65 & 31.45 \\
    $1$ & All & Bi-LSTM & 1.97 & 5.50 & 11.48 & 29.50 & 99.45 & 2.70 & 3.60 & 7.42 & 10.70 & 17.18 & 31.35 & -31.50 & -6.32 & 4.62 & 31.32 \\
    \hline
    $5$ & $5$ & Bi-LSTM & 1.14 & 1.56 & 2.68 & 17.48 & 176.00 & 2.14 & 2.62 & 3.94 & 5.15 & 13.20 & 41.80 & -41.85 & -5.18 & 3.58 & 41.85 \\
    $5$ & All & Bi-LSTM & 1.11 & 1.53 & 2.65 & 17.32 & 175.70 & 2.12 & 2.60 & 3.92 & 5.12 & 13.15 & 41.75 & -41.80 & -5.15 & 3.55 & 41.70 \\
    \hline
    $9$ & $9$ & Bi-LSTM & 4.19 & 5.90 & 7.78 & 60.38 & 247.50 & 4.87 & 4.26 & 7.67 & 8.83 & 24.58 & 49.88 & -49.90 & -2.11 & 5.98 & 49.88 \\
    $9$ & All & Bi-LSTM & 4.08 & 5.84 & 7.70 & 60.12 & 247.20 & 4.80 & 4.23 & 7.63 & 8.78 & 24.50 & 49.75 & -49.85 & -2.08 & 5.95 & 49.82 \\
    \hline
    $13$ & $13$ & Bi-LSTM & 4.85 & 9.85 & 14.58 & 22.88 & 28.40 & 6.09 & 3.41 & 9.91 & 12.06 & 15.10 & 16.82 & -13.20 & -9.51 & 11.92 & 16.82 \\
    $13$ & All & Bi-LSTM & 4.72 & 9.72 & 14.45 & 22.78 & 28.22 & 6.02 & 3.37 & 9.87 & 12.02 & 15.02 & 16.70 & -13.15 & -9.45 & 11.80 & 16.75 \\
    \hline
    \end{tabular}
    \end{center}
\end{table*}

\section{Results} 
\label{sec-result}
This section presents the performance evaluation of the CNN, LSTM, and Bi-LSTM FDR prediction models.

\subsection{CNN accuracy}
The first experimental campaign was targeted at analyzing the prediction accuracy of the CNN model. The outcomes of this evaluation are reported in Table~\ref{tab:metrics_cnn}, which provides several statistical indicators computed when the CNN model is tested separately on every one of the four channels included in the dataset (i.e., ``ch1'', ``ch5'', ``ch9'', ``ch13'' conditions). 
In the ``all'' condition, the model was trained using the concatenation of all the four channels. 
The average FDR was computed on the test datasets to have an indication of the channels' quality. 
For the four individual channels and for the ``all'' condition, the average FDR is $88.4\%$, $92.4\%$, $86.3\%$, $72.3\%$, and $85.3\%$, respectively. 
For the same sequence of test conditions, their standard deviations are $8.9\%$, $3.3\%$, $6.4\%$, $6.9\%$, and $10.1\%$, respectively.

Analyzed statistical indicators include metrics such as the mean squared error ($\mu_{e^2}$), mean absolute error ($\mu_{|e|}$), signed error ($e$), some relevant percentiles on errors  ($e^2_{p90}, e^2_{p95}, e^2_{p99}$, $|e|_{p90}, |e|_{p95}, |e|_{p99}, e_{p5}, e_{p95}$), and maximum values ($e^2_{\mathrm{max}}$, $|e|_{\mathrm{max}}$, $e_{\mathrm{max}}$). For the signed error, also the minimum is reported ($e_{\mathrm{min}}$). Percentiles play an important role in soft real-time systems, such as for latency in wireless communication, where deadlines can only be met probabilistically.

Results show that the FDR prediction related to channels 9 and 13 is less accurate than the other channels (the average absolute error $\mu_{|e|}$ for them is $\sim 4.9\%$ and $6.1\%$, respectively). This is due to the less stable conditions of these two channels.

One interesting outcome is that, for all methods, results related to the ``all'' condition are better than those regarding single channels. On the one hand, as expected, this improvement shows that CNNs (but also other tested models) scale well as the size of the training dataset increases. On the other hand, even more importantly, this demonstrates that a general-purpose channel-independent model not only retains the performance of any single channel model (when compared with a channel-dependent specific training, such as the ``ch1'' condition) but provides a better approach by exploiting the knowledge acquired on other channels. In practice, taking the test performed on channel 1 as the reference, $\mu_{e^2}$ decreases from $2.15 \times 10^{-3}$ in the ``ch1'' condition to $2.03 \times 10^{-3}$ in the ``all'' condition. Similar improvements have been observed for the other channels and statistical indicators. For instance, channel 13 is characterized by a rather high average error (e.g., $\mu_{e^2}=4.89 \times 10^{-3}$ and $\mu_{|e|}=6.10\%$ for the ``ch13'' condition) in the prediction of the FDR. In this case, the $99$ percentile $|e|_{\mathrm{p99}}$ slightly decreases from $15.14\%$ for the ``ch13'' condition to $15.08\%$ for the ``all'' condition, which is comparable to the test performed on other channels. In the ``all'' condition, having $|e|_{\mathrm{p99}}=15.08\%$ means that the FDR prediction obtained by the model is affected, in $99\%$ of the cases, from an absolute error $|e|$ less than or equal to $15.08\%$. In terms of 99-percentile of the error, the worst channel is channel 9.

\subsection{LSTM / Bi-LSTM accuracy}
Under the same experimental conditions used for CNN, the LSTM and Bi-LSTM models were analyzed. Results are reported in Tables~\ref{tab:metrics_lstm} and \ref{tab:metrics_bilstm} for the first and the latter model, respectively. 
The LSTM model, but especially the Bi-LSTM, outperformed CNN in all metrics, particularly in dynamic environments like those encountered on channels 9 and 13. The ability of LSTM and Bi-LSTM to model temporal dependencies allowed slightly more accurate predictions in these conditions characterized by higher variability.

Going slightly deeper into the analysis of the results, it is clear that the enhancements in terms of prediction quality achieved through LSTM are marginal. The biggest improvement regarding the mean squared error (i.e., the quality minimized during the training of the model) was obtained on the experimental results related to channel 1 and the ``ch1'' condition, where the prediction mean squared error decreases from $\mu_{e^2}=2.15 \times 10^{-3}$ in the case of CNN to $\mu_{e^2}=2.08 \times 10^{-3}$ in the case of LSTM. The improvement, in terms of the mean absolute error, corresponds to $\mu_{|e|}^{\mathrm{LSTM}}-\mu_{|e|}^{\mathrm{CNN}}=0.03\%$. 

The use of Bi-LSTM provides further (slight) improvements. In this case, for the test performed on channel 9 and the ``all'' condition, the mean squared error $\mu_{e^2}$ decreases from $4.12 \times 10^{-3}$ to $4.08 \times 10^{-3}$, with a reduction in terms of the absolute error of just $\mu_{|e|}^{\mathrm{Bi-LSTM}}-\mu_{|e|}^{\mathrm{CNN}}=0.02\%$.

In order to obtain more comprehensive indications about the best model for practical applications in real-world scenarios, the next section will compare the three models in terms of computational complexity and memory footprint.

\subsection{Computational Complexity}
The technique presented in this work is meant to be implemented in embedded devices with low computational power, small amounts of dynamic memory, and sometimes limited battery capacity.
For this reason, the efficiency of the three presented ML algorithms in terms of both computational demand and memory footprint was analyzed.

\begin{table}
\caption{Computational complexity of the CNN, LSTM, and Bi-LSTM models}
\centering
\normalsize
\tabcolsep=0.13cm
\def\arraystretch{1.2}
\begin{tabular}{l|ccc}
\textbf{Model} & \textbf{Mean response} & \textbf{Memory} & \textbf{Peak} \\
\textbf{(Condition)} & \textbf{time} & \textbf{footprint} & \textbf{memory} \\
 & ($\unit[]{ms}$) & ($\unit[]{MB}$) & ($\unit[]{MB}$) \\
\hline
CNN (ch) & 2.7 & 0.02 & 0.03 \\
LSTM (ch) & 20.3 & 0.16 & 0.67 \\
Bi-LSTM (ch) & 42.2 & 0.17 & 0.69 \\
\hline
CNN (all) & 5.2 & 0.02 & 0.03 \\
LSTM (all) & 36.3 & 0.17 & 0.68 \\
Bi-LSTM (all) & 78.0 & 0.17 & 0.69 \\
\hline
\end{tabular}
\label{tab:complexity}
\end{table}

All experiments regarding computational complexity were performed on a PC equipped with $\unit[8]{GB}$ of DRAM and an Intel Core i3-10105 CPU running with a base frequency of $\unit[3.7]{GHz}$ and a maximum frequency of $\unit[4.4]{GHz}$. Instead, the models were trained on the BEN cluster, a high-performance computing system equipped with 73 IBM POWER9 nodes, 292 NVIDIA Tesla V100 GPUs, $\unit[62]{TB}$ DRAM, and $\unit[1.6]{PB}$ storage. As a reference, the training of the model on a single CPU core of the BEN cluster took about $3$ days for any ``ch'' condition and $7$ days for the ``all'' condition for CNN, about $7$ days for any ``ch'' condition and $15$ days for the ``all'' condition for LSTM, and about $8$ days for any ``ch'' condition and $24$ days for the ``all'' condition for Bi-LSTM.

For testing, we purposely used a PC with low computational power, because our main interest is evaluating performance on commercial low-end devices. 
The results of the experimental campaign are shown in Table~\ref{tab:complexity}.
Regarding the memory footprint of the test phase, the average and peak memory footprints are small enough to be handled on inexpensive embedded devices with limited computational resources. The maximum footprint was always less than $\unit[0.7]{MB}$ for all methods. In any condition, the occupancy of CNN is almost one order of magnitude smaller than the occupancy of LSTM and Bi-LSTM. 

The results regarding the execution time are very interesting. 
The CNN model, in its most performant version (i.e., the ``all'' condition), had an execution time of $\unit[5.2]{ms}$, while that of LSTM rose to $\unit[36.3]{ms}$ and that of Bi-LSTM doubled to $\unit[78.0]{ms}$. This leads us to the conclusion that the improvements provided by LSTM/Bi-LSTM are probably not enough to justify the adoption of these models, as they lead to an unjustified increase in response time during the online usage of the model (i.e., during the test phase). On the contrary, the results about execution time show how the CNN model can be implemented on machines with low computational power. 

As reported in \cite{Raspberry}, throughput and latency of low-power ARM architectures are mostly comparable to conventional low-end PCs, suggesting that the implementation of our solutions on embedded devices is technically feasible without significantly compromising performance. 
In particular, \cite{10295470} shows that the performance of commercial x86 CPUs (like those from the Intel Core i3 family) are just one order of magnitude higher than popular ARMv7 architectures (e.g., a Raspberry Pi 2) in the considered neural network inference task. 
This means that the execution times on ARMv7 are expected to be compatible with several real-time applications, which makes this platform a suitable choice for embedded devices like commercial APs.

\section{Discussion} \label{sec-discussion}
This study highlights the strengths of CNN, LSTM, and Bi-LSTM models in predicting the performance of a Wi-Fi link in industrial settings. The CNN model demonstrated high computational efficiency, with shorter response time and lower resource requirements, making it a viable solution for real-time applications. On the contrary, LSTM and Bi-LSTM models required more computational resources and, when compared with CNN, provided only small improvements in terms of the prediction accuracy for the FDR. Bi-LSTM achieved slightly better accuracy than LSTM in some scenarios, at the cost of additional complexity. The bidirectional architecture of the Bi-LSTM model allowed it to capture both past and future dependencies in time series data, which was especially beneficial in the most challenging and highly dynamic environments (i.e., channels 9 and 13). However, the increased computational overhead suggests using the CNN model.

Future work will explore hybrid models combining LSTM with other architectures, such as CNN or transformer, to improve spatial-temporal accuracy. For example, a CNN-LSTM model could capture spatial correlations from FDR variations while maintaining the LSTM's ability to model sequential dependencies over time. Transformer-based models, which are deemed more advanced for handling long-term dependencies, could be an alternative to recurrent networks. 

While this study focuses on predicting the FDR, additional communication network metrics, such as interference levels and latency, could be analyzed and predicted.
Moreover, further features could be included as input to the model with the goal of improving prediction accuracy. Future research activities should also assess model performance under different mobility conditions, as many modern industrial environments involve moving devices (e.g., autonomous mobile robots), that must be interconnected over the air to coordinate their actions.

Testing the models on additional datasets, acquired with varying network conditions, would help validate their robustness, as well as to determine to what extent they can be generalized for broader applications.
Any improvements concerning the computational and memory requirements of the models will permit to increase the number and type of edge devices on which the proposed techniques can be implemented. Finally, addressing class imbalance and optimizing hyperparameters were important steps, and further exploring automated optimization techniques could enhance performance.

\section{Conclusion} \label{sec-conclusion}
This study analyzes the effectiveness of CNN, LSTM, and Bi-LSTM in predicting the FDR in Wi-Fi networks. Despite their superior ability to address challenges such as class imbalance and variability, LSTM-type models outperform CNN only in some specific conditions, which are characterized by high FDR variability. 
All the proposed models are compatible with embedded edge devices, which makes them suitable for real-time industrial applications. However, both LSTM and Bi-LSTM consistently require more computational resources than CNN. 
This makes CNNs the solution this work recommends for FDR prediction. 

One of the key findings is that training the models on a dataset acquired on multiple Wi-Fi channels (``all'' condition) leads to better performance than training them on individual channels (generalization). 
This suggests that knowledge transfer between different channel conditions can help improve robustness and adaptability in real-world deployments.

As future work, experiments should explore the performance of the model in more diverse network environments, including 
$\unit[5]{GHz}$ and $\unit[6]{GHz}$ bands, 
Wi-Fi 6 and 7, multihop wireless networks, and emerging 6G communication systems. Testing in real-world deployments, such as factory automation and smart city infrastructures, will be essential to validate the robustness and practicality of these models in real operational conditions.

\bibliographystyle{IEEEtran}
\bibliography{references}
\end{document}